\documentclass[conference]{IEEEtran}
\usepackage{amsmath,amsfonts}
\usepackage{algorithmic}
\usepackage{algorithm}
\usepackage{array}
\usepackage{textcomp}
\usepackage{stfloats}
\usepackage{url}
\usepackage{verbatim}
\usepackage{graphicx}
\usepackage{epstopdf}
\usepackage{subfig}
\usepackage{subfloat}
\usepackage{ragged2e}
\usepackage{cite}
\usepackage{tabularx}
\def\BibTeX{{\rm B\kern-.05em{\sc i\kern-.025em b}\kern-.08em
    T\kern-.1667em\lower.7ex\hbox{E}\kern-.125emX}}
\usepackage{balance}
\makeatletter

\newcommand{\Rmnum}[1]
{\expandafter\@slowromancap\romannumeral #1@}
\makeatletter
\begin{document}
\title{Event-Triggered Observer-Based Fixed-Time Consensus Control for Uncertain Nonlinear Multiagent Systems with Unknown States}
\author{\IEEEauthorblockN{1\textsuperscript{st} Kewei Zhou}
\IEEEauthorblockA{\textit{Southwest University} \\
\textit{Westa College}\\
Chongqing, China \\
swukeweizhou@163.com}
\and
\IEEEauthorblockN{2\textsuperscript{nd} Ziming Wang}
\IEEEauthorblockA{\textit{The Hong Kong University of Science and Technology (Guangzhou)} \\
\textit{Robotics and Autonomous Systems Thrust}\\
Guangzhou, China \\
wwwangziming@163.com}
\and
\IEEEauthorblockN{3\textsuperscript{rd} Zhihao Chen}
\IEEEauthorblockA{\textit{Wuhan University} \\
\textit{Electronic Information School}\\
Wuhan, China \\
czh011129@126.com}
\and
\IEEEauthorblockN{4\textsuperscript{th} Xin Wang}
\IEEEauthorblockA{\textit{Southwest University} \\
\textit{College of Electronics and Information Engineering}\\
Chongqing, China \\
xinwangswu@163.com}}
\maketitle
\begin{abstract}
This paper introduces a novel approach for achieving fixed-time tracking consensus control in multiagent systems (MASs). Departing from the reliance on traditional controllers, our innovative controller integrates modified tuning and Lyapunov functions to guarantee stability and convergence. Furthermore, we have implemented an event-triggered strategy aimed at reducing the frequency of updates, alongside an output-feedback observer to manage unmeasured states effectively. To address the challenges posed by unknown functions and algebraic-loop problems, we opted for radial basis function neural networks (RBF NNs), chosen for their superior performance. Our methodology successfully mitigates Zeno's behavior and ensures stability within a narrowly defined set. The efficacy of our proposed solution is validated through two illustrative simulation examples.
\end{abstract}


\section{Introduction}
The consensus control of multiagent systems (MASs), such as opinion dynamics, sensor networks, and stand-alone supply systems, have been extensively researched in academia over a long period due to their variety of applications \cite{ref1,ref2,ref5}. In our work, we mainly research the tracking consensus control for leader-follower MASs. The practical condition are more likely to increase the complexity of the signals when designing the controller for MASs. In \cite{ref10}, radial basis function neural networks (RBF NNs) have proved to be very useful in dealing with unknown and complex terms in nonlinear systems so they are widely used. So we can consider using RBF NNs to approximate the unknown uncertain terms that existed in system functions and the derivatives of Lyapunov functions. In addition, it is difficult to measure all state variables in practical systems because of the cost of the measuring equipment. Therefore, Chen and Li \cite{ref13} designed the observer-based distributed consensus controllers for second-order MASs. However, low-order MASs are also inapplicable to most physical systems. So Zhang et al \cite{ref14} established the fuzzy output-feedback observer for high-order MASs to make sure that the system states can be estimated during the set finite time. Due to their quicker convergence, and better robustness for systems, the finite time consensus control-based MASs are currently in widespread use, such as \cite{ref15,ref16}. However, the settling time of the finite-time consensus problem is related to the initial conditions in their works. For these obstacles, Huang \cite{ref23} and Wang \cite{ref26} constructed adaptive fixed-time tracking control. However, the fixed-time observer-based control scheme has not been considered widely in previous studies, which motivates our research.

In past studies, the time-sampling control strategy is used wildely and the signals are updated over time in the conventional work, such as \cite{ref14,ref23}. However, while this approach may appear to be energy intensive, it is important to consider that our focus is on the theoretical framework for reducing communication frequencies. Therefore, our research about an event-triggered mechanism was motivated in \cite{ref25,ref25.5,ref26.6,ref27,ref27.2,ref27.6} to reduce information transmission based on satisfying system performance. It is obvious that the event-triggered control has a better performance than time-triggered for MASs through the study of \cite{ref26.6}. In \cite{ref25}, an adaptive event-triggered output-feedback controller was designed for a class of uncertain nonlinear systems. Wang et al \cite{ref27} constructed a consensus control strategy of the nonlinear system based on fixed-time and event-triggered mechanisms. However, the state observer was not taken into account. Zhou and Wang\cite{ref25.5} utilize an event-triggered mechanism that takes into account unmeasured states, thereby enhancing the practicality of their control strategy. Our work focuses on a fixed-time event-triggered control strategy for MASs and guarantees that the convergence time is predetermined and independent of the system's initial conditions.

 Based on the above discussions, we suggest using a Lyapunov function, modified tuning functions, RBF NNs, state observer, and backstepping approach to create a strategy based on fixed-time and event-triggered participation. Finally, we make sure that every signal produced by the closed-loop systems has a uniform global bound. An overview of this article's primary contributions is provided below.\par

   \begin{enumerate}{}{}
\item{We developed a state observer to tackle the issue of unmeasured system states, a challenge often arising from the prohibitive costs or inherent limitations associated with measurement equipment.}
\item{This paper presents an innovative adaptive event-triggered control strategy for MASs with unobservable states. This approach effectively minimizes the frequency of controller activations, thereby optimizing the use of communication resources.}
\item{The traditional adaptive framework often encounters difficulties with error variables, challenging the assurance that the derivatives of the Lyapunov function align with stability criteria within a predetermined timeframe. To overcome this issue, we have devised a series of smooth functions specifically designed to secure fixed-time stability.}
\end{enumerate}

The main sections of our article are discussed in the following. Graph theory, system description, and RBF NNs will show in Section 2. Section 3 constructs the observer-based and event-triggered consensus controller. In Sections 4 and 5, the stability analysis and a practical example are presented. Lastly, the conclusion is shown in Section 6.

\section{problem description}

\subsection{Graph Theory}
 The directed graph $\breve{G}=(P,H,O)$ shows the $N$ followers' information communication, in which $P=\{p_{1},...,p_{N}\}$ denotes the node set, $H\subseteq P\times P $ is edge set, $O=[o_{ij}]\in R^{N\times N} $ is an adjacent matric. The edge from $j$ to $i$ is indicated by $(p_{j},p_{i})\in H $ and $o_{ij}>0$, or else, $o_{ij}=0$ and $ o_{ii}=0 $. The set of neighbors is indicated by $M_{i}=\{p_{j}|(p_{j},p_{i})\in H,i\neq j \}$. The in-degree matrix of $\breve{G}$ is represented by $T=diag\{t_{1},...,t_{N} \}$ with $t_{i}=\sum_{j\in N_{i}}o_{ij}$ . $L=T-O$ is the graph Laplacian matrix. In the leader-follower MASs, The augmented graph is denoted by $G^*=(P^*, H^*) $ with $P^*=\{p_{0},p_{1},...,p_{N}\}$ and $H^*\subseteq P^*\times P^*$, which contains all the followers. For a leader $p_{0}$, the adjacency matrix is represented by the diagonal matrix $C=diag\{c_{1},...,c_{N} \}$. As long as the leader is able to transmit information to follower $i$, there exists $c_{i}>0$; if not, $c_{i}=0$.\par
\subsection{System Description}
We describe the mathematical formulation of the system followers $i$ as:
\begin{align}
\dot{x}_{i,m}=& x_{i,m+1}+f_{i,m}(\bar{x}_{i,m}),1\leq m \leq n-1 \nonumber\\
\dot{x}_{i,n}=&\bar{u}_{i}+f_{i,n}(\bar{x}_{i,n}),i=1,2...N\nonumber\\
y_{i}=&x_{i,1}
\end{align}

in which $f_{i,m}(\cdot)$ are unknown smooth nonlinear functions. $\bar{x}_{i,m}=[x_{i,1},...,x_{i,m}]^T $  and $\bar{x}_{i,n}=[x_{i,1},...,x_{i,n}]^T$ denote the state variables. $y_{i}=x_{i,1}$ is system output signal and $\bar{u}_{i}$ is the input. $y_{0}$ is the output signal of the leader.

\subsection{RBF NNs}
 RBF NNs are widely used to estimate unknown continuous functions in the following:
\begin{align}
f(Z)=\aleph^T\ell(Z)
\end{align}
 where $Z$ stands for the input vector, $\ell(Z)=[\ell_{1}(Z),...,\ell_{m}(Z)]^T$ represents the basis vector with neuron number $m$ and $m>1$. $ \aleph=[\aleph_{1},...,\aleph_{m}]^T$ is the weights vector. When estimating the unknown smooth nonlinear function with RBF NNs, one obtains:
\begin{align}
f(Z)=\aleph^{*T}\ell(Z)+\psi(Z)
\end{align}
 where $\psi(Z)$ is the approximate error, and $a$ is a positive constant having $\parallel \psi(Z) \parallel<a$ . Then the optimal weight matrix $\aleph^{*}$  can be represented as follows:
\begin{align}
\aleph^* =\arg \min_{\aleph\in \mathbb{R}^m}\{ \sup_{Z \in \Omega_{Z}}|f(Z)-\aleph^T\ell(Z)| \}
\end{align}
 where $ \aleph^*=[\aleph^*_{1},...,\aleph^*_{m}]^T$, $\mathbb{R}^m$ and $\Omega_{Z}$ are compact sets of $\aleph^*$, $Z$. The Gaussian function $\ell_{i}(Z)$ as follows:
\begin{align}
\ell_{i}(Z)=\exp\left[\frac{-(Z-\iota_{i})^T(Z-\iota_{i})}{\varpi^2_{i}}\right],
\end{align}
 where $\iota_{i}=[\iota_{i,1},...,\iota_{i,m}]^T$ represents the centre vector and $\varpi_{i}$ indicates the width of the Gaussian function.

 \emph{Lemma \textsl{1\cite{ref34}:}}
For $\forall a\in \mathbb{R}$, the inequality is satisfied:
\begin{align}
0\leq\mid a \mid- a \, \tanh\left(\frac{a}{\lambda}\right) \leq \lambda\tau, \forall \lambda>0
\end{align}
 with $\tau=0.2785$.\par
 \emph{Lemma \textsl{2\cite{ref35}:}}
If there are nonlinear multiagent systems (1), considering the derivative of Lyapunov function $V(\zeta)$, one obtains:
\begin{align}
\dot{V}(\zeta)\leq -\bar{a}V^{\breve{\alpha}}(\zeta)-\bar{b}V^{\breve{\beta}}(\zeta)+\bar{c}
\end{align}
 where $\bar{a},\bar{b}>0$, $\breve{\alpha}\in(0,1)$, $\breve{\beta}>1$, and $\bar{c}< min \{(1-p)\bar{a},(1-p)\bar{b}\} (p\in(0,1))$. The fixed-time stability for MASs could be accomplished and the settling time $T_{0}$ has:
\begin{align}
T_{0}\leq T_{max}=\frac{1}{p \bar{a}(1-\breve{\alpha})}+\frac{1}{p \bar{b}(\breve{\beta}-1)}.
\end{align}\par
$Assumption\quad1:$ The leader's output signal has n-order continuous derivatives, and they are bounded and smooth.\par

\section{adaptive controller design}
 We create a state observer in the following to approximate the unknown states:
\begin{align}
\overset{.}{\hat{x}} _{i,m}=&{\hat{x}} _{i,m+1}-\mu_{i,m}(y_{i}-\hat{y}_{i}), 1\leq m\leq n-1 \nonumber\\
\overset{.}{\hat{x}} _{i,n}=&\bar{u} _{i}-\mu_{i,n}(y_{i}-\hat{y}_{i}) ,i=1,2,...,N
\end{align}
in which ${\hat{x}} _{i,m}$, $\hat{y}_{i}$ are the estimated value of $x_{i,m}$, $y_{i}$ and $\mu_{i,m}$ are the observer gains. The state observer is based on each follower's output signals inspired by \cite{ref14}.

The observer error is $e_{i,m}=x_{i,m}-{\hat{x}} _{i,m}$ for $1\leq i\leq N,1\leq m\leq n$.

And then
\begin{align}
\dot{e}_{i}=\chi_{i}\bar{e}_{i}+f_{i}(\bar{x}_{i,n})
\end{align}
where $
\bar{e}_{i}=[e_{i,1},...,e_{i,n}]^T,
\chi_{i}=\chi_{0}+M_{i}P,
\chi_{0}=\begin{bmatrix}
 0 &\bar{I}_{n-1} \\
0  &0
\end{bmatrix},
M_{i}=[\mu_{i,1},...,\mu_{i,n}]^T,
P=[1,0,...,0],\\ $ and $
f_{i}(\bar{x}_{i,n})=[f_{i}(x_{i,1}),...,f_{i}(x_{i,n})]^T.$

Choose $M_{i}$ such that $\chi_{i}=\chi_{0}+M_{i}P$ becomes a Hurwitz matrix. For any $\Re>0$, there exists a matrix $H_{i}=H^T_{i}>0$ that satisfies:
\begin{align}
\chi^T_{i}H_{i}+H_{i}\chi_{i}=-\Re \bar{I}
\end{align}
 where $\bar{I}$ is unit matrix.

We employ RBF NNs to solve the uncertainty and complexity of the uncertain nonlinear function, which yields:
\begin{align}
f_{i}(\bar{x} _{i,n})=\theta^T _{i}\breve{S}_{i}(\bar{x} _{i,n})+\psi _{i}(\bar{x} _{i,n})
\end{align}
where $\psi_{i}(\bar{x} _{i,n}) $ represents the approximate error satisfying a given precision level $\epsilon_{i}\geq \left \| \psi _{i}(\bar{x} _{i,n})  \right \| $.\\
with
\begin{align}
\theta _{i}=&diag\left [ \theta _{i,1},...,  \theta _{i,n}\right ]\nonumber\\
 \breve{S}_{i}(\bar{x} _{i,n})=&[\breve{S}_{i,1}(x _{i,1}),...,\breve{S}_{i,n}(\bar{x} _{i,n})]^T\nonumber\\
 \psi_{i}(\bar{x} _{i,n})=&[\psi_{i,1}(x _{i,1}),...,\psi_{i,n}(\bar{x} _{i,n})]^T\nonumber\\
 \epsilon_{i}=&\sqrt{\epsilon^2 _{i,1}+...+\epsilon^2 _{i,n}}\nonumber.
\end{align}

 Then, we have:
\begin{align}
\dot{e}_{i}=\chi_{i}e_{i}+\theta^T _{i}\breve{S}_{i}(\bar{x }_{i,n})+\psi _{i}(\bar{x }_{i,n}).
\end{align}

 We will select the Lyapunov function $V_{i,e}$ to analyze the stability and convergence of (13):
\begin{align}
V_{i,e}=e^T_{i}\Upsilon_{i}e_{i}
\end{align}
 in which $\Upsilon_{i}$ is a symmetric positive matrix and we define it at once. According to (13) and (14), one obtains:
\begin{align}
\dot{V}_{i,e} =e^T_{i}(\Upsilon_{i}\chi _{i}+\chi^T _{i}\Upsilon_{i}) e_{i}+2e^T_{i}\Upsilon_{i}\left[ \theta^T _{i}\breve{S}_{i}(\bar{x} _{i,n})+\psi _{i}(\bar{x} _{i,n}) \right].
\end{align}

Select $\eta _{i}=max\left \{ \left \| \theta _{i} \right \| ,\left \| \theta _{i,k} \right \|  \right \} $ with $ \left \| \theta _{i} \right \|\le \eta _{i}$ .According to Young's equality and $0<\breve{S}^T_{i}\breve{S}_{i}<1$. We have:
\begin{align}
2e^T_{i}\Upsilon_{i}\left [ \theta^T _{i}\breve{S}_{i}(\bar{x} _{i,n})+\psi _{i}(\bar{x} _{i,n}) \right ]\nonumber \\ \leq 2\parallel e_{i}\parallel^2 +\parallel \Upsilon_{i}\parallel^2\eta_{i}+\parallel \Upsilon_{i}\parallel^2\epsilon_{i}.
\end{align}

Combining (15) and (16), it has
\begin{align}
\dot{V}_{i,e}\leq e^T_{i}(\Upsilon_{i}\chi_{i}+\chi^T_{i}\Upsilon_{i}+2\bar{I})e_{i}+q_{0}
\end{align}
with $q_{0}=\parallel \Upsilon_{i}\parallel^2\eta_{i}+\parallel \Upsilon_{i}\parallel^2\epsilon_{i}$.

According (11) and (17), we know that the observer error satisfies $\dot{ V}(e)\leq -\breve{a}V^\alpha(e)+\breve{b} $ if $\Re_{i}>2$ . In which $\breve{a},\breve{b}$ are positive constants so that error is bounded ultimately.

\subsection{Control Design}
According to \cite{ref23}, we will introduce a class of smooth functions before the backstepping design.
\begin{align}
sg_{i,m}(\gamma_{i,m})=&\begin{cases}\frac{\gamma_{i,m}}{\left | \gamma _{i,m} \right | }, \left | \gamma _{i,m}\right | \ge \kappa _{i,m} \\\frac{\gamma_{i,m}}{(\kappa^2 _{i,m}-\gamma^2_{i,m})+\left | \gamma_{i,m} \right | } ,\left | \gamma _{i,m}\right | < \kappa _{i,m} \end{cases}\\
f_{i,m}=&\begin{cases}1, \left | \gamma _{i,m}\right | \ge \kappa _{i,m} \\0 ,\left | \gamma _{i,m}\right | < \kappa _{i,m} \end{cases}.
\end{align}
where $\kappa _{i,m}$ is a positive constant.

Combining (18) and (19), we have
\begin{align}
sg_{i,m}(\gamma_{i,m})\times f_{i,m}=\begin{cases}\frac{\gamma_{i,m}}{\left | \gamma _{i,m} \right | }, \left | \gamma _{i,m}\right | \ge \kappa _{i,m} \\0 ,\left | \gamma _{i,m}\right | < \kappa _{i,m} \end{cases}.
\end{align}

{\bf{Remark 1:}} An improved switched tuning function is proposed to avoid the potential singularity of the MASs. Comparing the controlled MASs in \cite{ref23}, we know that the chattering issue can be avoided by introducing $sat$ function.

Now we will start the controller design as follows.

Firstly, we introduce the error vector of the $i$ agent as follows:
\begin{align}
\gamma _{i,1}=&\sum_{j\in N} \Gamma _{ij}(y_{i}-y _{j})+t_{i}(y_{i}-y_{0})\nonumber\\
\gamma _{i,m}=&\hat{x}_{i,m}-\alpha _{i,m-1}.
\end{align}

{\it{Step 1}}: Based on backstepping technique, the Lyapunov function $V_{i,1}$ is selected in the following:
\begin{align}
V_{i,1}=\frac{1}{2}(\left | \gamma _{i,1}\right |-\kappa _{i,1}  )^2f_{i,1}+\frac{1}{2r_{i,1}}\tilde{\varphi}^T  _{i,1} \tilde{\varphi} _{i,1},
\end{align}
where $r_{i,1}$, $\varphi _{i,1}$ are positive constant. We can obtain the derivative of $\gamma_{i,1}$ by (21): $\dot{\gamma} _{i,1}
=s_{i}x_{i,2}+s_{i}f_{i,1}(x_{i,1})-\sum_{j\in N}\Gamma _{ij}[x_{j,2}+f_{j,1}(x_{j,1})]-t_{i}\dot{y}_{0}$ with $s_{i}=\sum_{j\in N}\Gamma _{ij}+t_{i}$.

According to (22) and the derivative of $\gamma_{i,1}$:
\begin{align}
\dot{V}_{i,1}=A_{i,1}[s_{i}\alpha _{i,1}+s_{i}e_{i,2}+s_{i}\gamma _{i,2}+F_{i,1}(Z_{i,1} )]-\frac{1}{r_{i,1}}\tilde{\varphi}^T  _{i,1} \dot{\hat{\varphi}}  _{i,1},
\end{align}
where $A_{i,1}=(\left | \gamma _{i,1} \right |-\kappa _{i,1} )sg_{i,1}( \gamma _{i,1} )f_{i,1}$, and $F(Z_{i,1})=s_{i}f_{i,1}(x_{i,1})-\sum_{j\in N}\Gamma _{ij}\left [ x_{j,2}+f_{j,1}(x_{j,1}) \right ]-t_{i}\dot{y}_{0}$ with $Z_{i,1}=\left [ x_{i,1},x_{j,1},x_{j,2},y_{0},\dot{y} _{0} \right ] ^T$.

{\bf{Remark 2:}} $F(Z_{i,1})$ cannot be applied in the control scheme, so $W_{i,1}^T\Xi_{i,1}(Z_{i,1})$ is used to approximate the unknown nonlinear terms $F(Z_{i,1})$. For $\Delta_{i,1}>0$, positive constant $\varphi _{i,1}=max\{\parallel W_{i,1}\parallel,\Delta_{i,1} \}$ and $\hat{\varphi} _{i,1}$ is estimation value of $\varphi_{i,1}$ whose error can be described as $\tilde{\varphi} _{i,1}=\varphi _{i,1} -\hat{\varphi} _{i,1}$.

$F(Z_{i,1})$ can be represented through RBF NNs:
\begin{align}
F(Z_{i,1})=&W_{i,1}^T\Xi_{i,1}(Z_{i,1})+\Lambda _{i,1}(Z_{i,1})\nonumber\\
\left | \Lambda _{i,1}(Z_{i,1}) \right | \le& \Delta _{i,1}
\end{align}
where $ \Lambda _{i,1}(Z_{i,1})$ denotes the approximate error.

Combining Lemma 1 and (24), we get
\begin{align}
A_{i,1}F_{i,1}(Z_{i,1}) \le&
 A_{i,1}[\left \| W_{i,1}^T \right \| \left \| \Xi _{i,1} \right \| +\Delta  _{i,1}] \nonumber\\\le&
  A_{i,1}\varphi _{i,1}g_{i,1} tanh(\frac{A_{i,1}g_{i,1}}{\bar{m}} )+\bar{m}\varsigma \varphi _{i,1}
\end{align}
where $\varphi _{i,1}=max\left \{ \left \| W_{i,1} \right \|,\Delta _{i,1}  \right \}$, $\bar{m}>0$ and $ g_{i,1}=1+\left \| \Xi _{i,1} \right \| $.

According to Young's inequality, we have
\begin{align}
A_{i,1}s_{i}e_{i,2}\le& e^T_{i}e_{i}+\frac{1}{4}(\left | \gamma _{i,1} \right |-\kappa _{i,1} )^2sg_{i,1}^2( \gamma _{i,1} )f_{i,1}^2s^2_{i}.
\end{align}

Based on (23)-(26), we proposed the virtual control scheme and the parameter adaptive law in the following:
\begin{align}
\alpha _{i,1}=-&\frac{1}{s_{i}} [a_{i,1}(\left | \gamma _{i,1} \right |-\kappa _{i,1} )^{2p-1}sg_{i,1}(\gamma _{i,1})+b_{i,1}\left | \gamma _{i,1} \right |\nonumber\\-&\kappa _{i,1} )^{2q-1}sg_{i,1}(\gamma _{i,1})+\frac{1}{4}A_{i,1}s^2_{i}+\hat{\varphi} ^T _{i,1}g_{i,1}tanh(\frac{A_{i,1}g_{i,1}}{\bar{m}} ) \nonumber\\ +&\frac{1}{2} (\left | \gamma _{i,1} \right |-\kappa _{i,1} )sg_{i,1}( \gamma _{i,1} )s_{i} +(\kappa _{i,2}+1)sg_{i,1}( \gamma _{i,1} )s_{i}],
\\
\dot{\hat{\varphi}} _{i,1}=&r_{i,1}A_{i,1}g_{i,1} tanh(\frac{A_{i,1}g_{i,1}}{\bar{m}})-\rho _{i,1}\hat{\varphi}  _{i,1},
\end{align}
where $a_{i,1}$, $b_{i,1}$ and $\rho _{i,1}$ are both designed positive parameters.

Invoking (24)-(28) into (23), and one obtains:
\begin{align}
\dot{V} _{i,1}\le -&a_{i,1}(\left | \gamma _{i,1} \right |-\kappa _{i,1} )^{2p}f_{i,1}-b_{i,1}(\left | \gamma _{i,1} \right |-\kappa _{i,1} )^{2q}f_{i,1}
\nonumber\\-&\frac{1}{2}s_{i}(\left | \gamma _{i,1} \right |-\kappa _{i,1} )^2f_{i,1}+e^T_{i}e_{i}+\frac{\rho _{i,1}}{r_{i,1}}\tilde{\varphi}^T _{i,1} \hat{\varphi} _{i,1}\nonumber\\+&\bar{m}\varsigma \varphi _{i,1}+s_{i}(\left | \gamma _{i,1} \right |-\kappa _{i,1})f_{i,1}[\left | \gamma _{i,1} \right |-(\kappa _{i,1}+1)].
\end{align}

{\it{Step m}}: The Lyapunov function $V_{i,m}$ is selected as:
\begin{align}
V_{i,m}=\frac{1}{2}(\left | \gamma _{i,m} \right |-\kappa _{i,m})^2f_{i,m}+\frac{1}{2r_{i,1}}\tilde{\varphi } ^T_{i,m} \tilde{{\varphi } } _{i,m}+V_{i,m-1}.
\end{align}

Take the derivative of $V_{i,m}$:
\begin{align}
\dot{V} _{i,m}=A_{i,m}\dot{\gamma}_{i,m}-\frac{1}{r_{i,m}} \tilde{\varphi}^T  _{i,m}\dot{\hat{\varphi}} _{i,m}-\dot{V} _{i,m-1},
\end{align}
where $A_{i,m}=(\left | \gamma _{i,m} \right |-\kappa _{i,m})sg_{i,m}(\gamma_{i,m})f_{i,m}$ and $\dot{\gamma} _{i,m}=\gamma _{i,m+1}+\alpha _{i,m}+\mu_{i,m}(y_{i}-\hat{y} _{i})-\dot{\alpha}  _{i,m-1}$ , then we obtain:
\begin{align}
\dot{V}_{i,m}=&A_{i,m}[\gamma _{i,m+1}+\alpha _{i,m}+\mu_{i,m}(y_{i}-\hat{y} _{i})-\dot{\alpha}  _{i,m-1}]\nonumber\\-&\frac{1}{r_{i,m}} \tilde{\varphi}^T  _{i,m}\dot{\hat{\varphi}} _{i,m}-\dot{V} _{i,m-1}
\end{align}
where $\dot{\alpha}  _{i,m-1} =\frac{\partial \alpha _{i,m-1}}{\partial x_{i,1}}(e_{i,2}+\hat{x} _{i,2}+f_{i,1}(x_{i,1})) + \sum_{j\in N}  \frac{\partial \alpha _{i,m-1}}{\partial \hat{x}_{j,1}}(e_{j,2}+\hat{x} _{j,2}+f_{j,1}(x_{j,1})) + \sum_{q=1}^{m-1}  \frac{\partial \alpha _{i,m-1}}{\partial \hat{\varphi }_{i,q} } \dot{\hat{\varphi }}  _{i,q} +\sum_{q=0}^{m-1}\frac{\partial \alpha _{i,m-1}}{\partial y_{0}^{(q)}} y^{q+1}_{0}+ \sum_{q=2}^{m-1} \sum_{j\in N}\frac{\partial \alpha _{i,m-1}}{\partial \hat{x} _{j,q}}(\hat{x} _{j,q+1}+\mu_{j,q}(y_{i}-\hat{y} _{i})),$ and $F_{i,m}(\nabla_{i,m})=-[\frac{\partial \alpha _{i,m-1}}{\partial x_{i,1}}(\hat{x} _{i,2}+f_{i,1}(x_{i,1})) + \sum_{j\in N}  \frac{\partial \alpha _{i,m-1}}{\partial \hat{x}_{j,1}}(\hat{x} _{j,2}+f_{j,1}(x_{j,1})) + \sum_{q=1}^{m-1} \frac{\partial \alpha _{i,1}}{\partial \hat{\varphi }_{i,q} } \dot{\hat{\varphi }}  _{i,q} +\sum_{q=0}^{m-1}\frac{\partial \alpha _{i,m-1}}{\partial y_{0}^{(q)}} y^{q+1}_{0}+ \sum_{q=2}^{m}\sum_{j\in N}\frac{\partial \alpha _{i,m-1}}{\partial \hat{x} _{j,q}}\hat{x} _{j,q+1} +A_{i,m}^2[\mu^2_{i,m}+(\frac{\partial \alpha _{i,m-1}}{\partial x_{i,1}})^2+(\sum_{j\in N}\frac{\partial \alpha _{i,m-1}}{\partial \hat{x}_{j,1}})^2+(\sum_{q=2}^{m}\sum_{j\in N} \frac{\partial \alpha _{i,m-1}}{\partial \hat{x} _{j,q}} \mu_{j,q})^2]$.

 According to (31) and the derivation of $\alpha_{i,m-1}$, one obtains:
\begin{align}
\dot{V}_{i,m}=&A_{i,m}[\gamma _{i,m+1}+\alpha _{i,m}+F_{i,m}(Z_{i,m})]-\frac{1}{r_{i,m}} \tilde{\varphi}^T  _{i,m}\dot{\hat{\varphi}} _{i,m}\nonumber\\-&\dot{V} _{i,m-1}
\end{align}

We employing RBF NNs to estimate $F_{i,m}(Z_{i,m})$ as follows:
\begin{align}
F(Z_{i,m})=&W_{i,m}^T\Xi _{i,m}(Z_{i,m})+\Lambda _{i,m}(Z_{i,m})\nonumber\\ \left | \Lambda _{i,m}(Z_{i,m}) \right | \le& \Delta _{i,m}.
\end{align}

According to Lemma 1, we get:
\begin{align}
A_{i,m}F_{i,m}(Z_{i,m})\le&
 A_{i,m}[\left \| W_{i,m}^T \right \| \left \| \Xi _{i,m} \right \| +\Delta  _{i,m}]\nonumber\\ \le&
  A_{i,m}\varphi _{i,m}g_{i,m} tanh(\frac{A_{i,m}g_{i,m}}{\bar{m}} )+\bar{m}\varsigma \varphi _{i,m}
\end{align}
where $\varphi _{i,m}=max\left \{ \left \| W_{i,m} \right \|,\Delta _{i,m}  \right \}$ and $ g_{i,m}=1+\left \| \Xi _{i,m} \right \| $.

Then we propose the virtual control scheme and the parameter adaptive law:
\begin{align}
\alpha _{i,m}=-&[a_{i,m}(\left | \gamma _{i,m} \right |-\kappa _{i,m} )^{2p-1}sg_{i,m}(\gamma _{i,m})\nonumber\\+&b_{i,m}\left | \gamma _{i,m} \right |-\kappa _{i,m} )^{2q-1}sg_{i,m}(\gamma _{i,m})\nonumber\\+&\hat{\varphi} ^T _{i,m}g_{i,m}tanh(\frac{A_{i,m}g_{i,m}}{\bar{m}} ) \nonumber\\+&(\left | \gamma _{i,m} \right |-\kappa _{i,m} )sg_{i,m}( \gamma _{i,m} )\nonumber\\+&(\kappa _{i,m+1}+1)sg_{i,m}(\gamma _{i,m})]\\
\dot{\hat{\varphi}} _{i,m}=&r_{i,m}A_{i,m}g_{i,m}tanh(\frac{A_{i,m}g_{i,m}}{\bar{m}})-\rho _{i,m}\hat{\varphi}_{i,m}.
\end{align}

Combining (34)-(37) into (33), one obtains:
\begin{align}
V_{i,m} \le -&\sum_{k=1}^{m}  a_{i,k}(\left | \gamma _{i,k} \right |-\kappa _{i,k} )^{2p}f_{i,k}\nonumber\\-&\sum_{k=1}^{m}  b_{i,k}(\left | \gamma _{i,k} \right |-\kappa _{i,k} )^{2q}f_{i,k}+me^T_{i}e_{i}\nonumber\\+&\sum_{k=1}^{m}\frac{\rho _{i,k}}{r_{i,k}}\tilde{\varphi}  ^T_{i,k}\hat{\varphi }_{i,k}+\sum_{k=1}^{m}\bar{m}\varsigma \varphi _{i,k}\nonumber\\-&[\left | \gamma_{i,m+1} \right |- (\kappa _{i,m+1}+1)](\left | \gamma_{i,m} \right | -\kappa _{i,m})f_{i,m}+\hbar_{i,m}
\end{align}
where $\hbar _{i,m}= -\frac{1}{2}(\left | \gamma _{i,m-1} \right | - \kappa _{i,m-1})^2 f_{i,m-1} +(\left | \gamma _{i,m-1} \right | - \kappa _{i,m-1}) f_{i,m-1}[\left | \gamma _{i,m} \right |-(\kappa _{i,m}+1) ] -\frac{1}{2}(\left | \gamma _{i,m} \right | -\kappa _{i,m})^2 f_{i,m}  $. It is apparent that we get $\hbar _{i,m} <0$.

{\it{Step n}}: The last updated scheme of the agent and the controller are designed in this step. Construct the Lyapunov function $V_{i,n}$ as
\begin{align}
V_{i,n}=\frac{1}{2}(\left | \gamma _{i,n} \right |-\kappa _{i,n})^2f_{i,n}+\frac{1}{2r_{i,n}}\tilde{\varphi } ^T_{i,n} \tilde{{\varphi } } _{i,n}+V_{i,n-1}.
\end{align}

Differentiating $V_{i,n}$ shows that:
\begin{align}
\dot{V} _{i,n}=A_{i,n}\dot{\gamma}_{i,m}-\frac{1}{r_{i,n}} \tilde{\varphi}^T  _{i,n}\dot{\hat{\varphi}} _{i,n}-\dot{V} _{i,n-1}
\end{align}
where $\dot{\gamma} _{i,n} =u_{i}+\mu_{i,n}(y_{i}-\hat{y}_{i})-\dot{\alpha}_{i,n-1}$ and $A_{i,n}=(\left | \gamma _{i,n} \right |-\kappa _{i,n})sg_{i,n}(\gamma_{i,n})f_{i,n}$, then one obtains:
\begin{align}
\dot{V}_{i,n}=A_{i,n}[u_{i}+F_{i,n}(Z_{i,n})]-\frac{1}{r_{i,n}} \tilde{\varphi}^T  _{i,n}\dot{\hat{\varphi}} _{i,n}-\dot{V} _{i,n-1}.
\end{align}

Similar to the RBF NNs that we discussed in {\it{step m}}, one gets:
\begin{align}
F(Z_{i,n})=&W_{i,n}^T\Xi _{i,n}(Z_{i,n})+\Lambda _{i,n}(Z_{i,n})\nonumber\\ \left | \Lambda _{i,n}(Z_{i,n}) \right | \le& \Delta _{i,n}.
\end{align}

{\bf{Remark 3:}} In \cite{ref40}, "explosion of complexity" was eliminated with the proposal of the command filter but they need to compensate for the signal due to the errors. $\sum_{q=0}^{n-1}\frac{\psi \alpha _{i,n-1}}{\psi y_{0}^{(q)}} y^{q+1}_{0}$ is directly approximately by NNs to resolve the issue in our article. Therefore, we can avoid constructing and compensating error systems in \cite{ref23,ref40}.

\subsection{Event-triggered Controller}

The event-triggered tracking consensus controller is constructed in the following:
\begin{align}
w_{i}(t)=\alpha _{i,n}-\xi _{i}tanh(\frac{A_{i,n}\xi _{i}}{\varepsilon _{i}}).
\end{align}

 Invoking the same consideration with Step m, and $\alpha _{i,n}$ is denoted as $\alpha _{i,n}=-[a_{i,n}( \left | \gamma _{i,n}   \right | -\kappa _{i,n} )^{2p-1}sg_{i,n}(\gamma _{i,n})+b_{i,n} (| \gamma _{i,n} |-\kappa _{i,n} )^{2q-1}sg_{i,n}(\gamma _{i,n})+\frac{1}{2}( |\gamma _{i,n} |-\kappa _{i,n} )sg_{i,n}( \gamma _{i,n} )+\\ \hat{\varphi} ^T _{i,n}g_{i,n} tanh(\frac{A_{i,n}g_{i,n}}{\bar{m}} )]$. The adaptive law is considered:\\$
\dot{\hat{\varphi}} _{i,n}=r_{i,n}A_{i,n}g_{i,n}tanh(\frac{A_{i,n}g_{i,n}}{\bar{m}})-\rho _{i,n}\hat{\varphi}_{i,n}
$.

And the triggering event is constructed in the following:
\begin{align}
\bar{u}_{i}(t)=&w_{i}(t_{\gamma }) ,\forall t\in [t_{\gamma },t_{\gamma +1})\\
t_{\gamma +1}=&inf\left \{ t\in R^N||\breve{\phi }_{i}(t)\ge \xi ^*_{i} \right \} ,t_{1}=0
\end{align}
where $\varepsilon_{i}$, $\xi_{i}$ , $\xi^*_{i}$ and $\xi_{i}>\xi^*_{i}$ are designed constant. $\breve{\phi} _{i}(t)=w_{i}(t)-\bar{u}_{i}(t)$ is the event's measurement error. The system will respond when event meets certain conditions. When (45) is triggered, the time will be labeled as $t_{\gamma}$ and $\bar{u}_{i}(t_{\gamma+1})$ will update the controller. During the time $t\in [t_{\gamma },t_{\gamma +1})$, the system signal keeps a constant. $w_{i}(t_{\gamma})$. Therefore, the continuous time-varying parameter  $\breve{\phi} ^*_{i}(t)$ fulfill:
$\breve{\phi} ^*_{i}(t_{\gamma})=0$ and $\breve{\phi} ^*_{i}(t_{\gamma+1})=\pm 1$ with $\mid \breve{\phi} ^*_{i}(t)\mid\leq 1$. We get $w_{i}(t)=\bar{u}_{i}(t)+\breve{\phi} _{i}(t)\xi ^*_{i}$.

In light of Lemma 1, we get:
\begin{align}
-\breve{\phi} ^*_{i}(t)\xi ^*_{i}A_{i,n}-\xi _{i}A_{i,n}\xi _{i}tanh(\frac{A_{i,n}\xi _{i}}{\varepsilon _{i}})\le 0.2875\varepsilon _{i}
\end{align}

Differentiating $V_{i,n}$ based on (39)-(46), we get:
\begin{align}
\dot{V}_{i,n} \le\sum_{k=1}^{n} \{- a_{i,k}(\left | \gamma _{i,k} \right |-\kappa _{i,`1k} )^{2p}f_{i,k}+\frac{\rho _{i,k}}{r_{i,k}}\tilde{\varphi}  ^T_{i,k}\hat{\varphi }_{i,k}\nonumber\\-b_{i,k}(\left | \gamma _{i,k} \right |-\kappa _{i,k} )^{2q}f_{i,k}+\bar{m}\varsigma \varphi _{i,k}\}+ne^T_{i}e_{i}+0.2875\varepsilon _{i}+\hbar_{i,n}
\end{align}
where $\hbar_{i,n}$ is similar to {\it{step m}} and $\hbar_{i,n}<0$.

{\bf{Remark 4:}} In this paper, we have designed the adaptive controllers in (27), and (36) and the control update schemes in (28), and (37) to achieve the stability of Lyapunov in fixed time. In addition, in order to save resources, we also consider event triggering and design an adaptive controller in (43) based on conditional triggering so that the controller is triggered when (45) is satisfied.

\section{STABLITY ANALYSIS}
We constructed state observers and established event-triggered tracking consensus controllers based on the previous discussion. Now we summarize the acquired results as follows.\par
$Assumption\quad2:$ There exists the symmetric matrix  $\Upsilon_{i}>0$ and $\chi _{i}$ satisfying $\Upsilon _{i}\chi _{i}+\chi ^T _{i}\Upsilon _{i}+(\Theta _{i}+(2+n))I<0$.
{\bf{Theorem:}}
For MASs(1) and event-triggered controller(43), we can get
  \begin{enumerate}
\item{The system's closed-loop signals must be bounded completely.}
\item{The tracking consensus errors from the followers' output signal to the leader's reference signal can approach a region in a fixed-time $T_{0}$, namely,\\$\lim_{t \to T_{0}}\left | \gamma _{i,1} \right |\le \kappa _{i,1} $. And the system error signals $\tilde{\varphi}_{i,m}$ and $\gamma_{i,m}$ are ultimately bounded.}
\end{enumerate}

{\bf{Proof:}} The whole Lyapunov function $V$ can be discussed in the following:
\begin{align}
V=\sum_{i=1}^{N} \sum_{k=1}^{n} V_{i,k}+\sum_{i=1}^{N} V_{i,e}.
\end{align}

Take the derivative of $V$ :
\begin{align}
\dot{V}\le &\sum_{i=1}^{N} \{ e^T_{i}(\Upsilon _{i}\chi _{i}+\chi ^T _{i}\Upsilon _{i}+(\Theta _{i}+2)I)e_{i}\nonumber\\-&\sum_{k=1}^{n}  a_{i,k}(\left | \gamma _{i,k} \right |-\kappa _{i,k} )^{2p}f_{i,k}-\sum_{k=1}^{n}  b_{i,k}(\left | \gamma _{i,k} \right |-\kappa _{i,k} )^{2q}f_{i,k}\nonumber\\+&\sum_{k=1}^{n}\frac{\rho _{i,k}}{r_{i,k}}\tilde{\varphi}^T_{i,k}\hat{\varphi }_{i,k}+\sum_{k=1}^{n}\bar{m}\varsigma \varphi _{i,k}+0.2875\varepsilon _{i}+q_{0} \}
\end{align}
where we define $\varpi_{i}=([\Theta_{i}-(2+n)]/[\bar{\lambda}_{max}(\Upsilon_{i})])$ with $\bar{\lambda}_{max}(\Upsilon_{i})$ being the maximum eigenvalue of  $\Upsilon_{i}$.

Combining (49), Lemma 1 in \cite{ref39} and Young's inequality, we get
\begin{align}
\dot{V}\le& \sum_{i=1}^{N} \sum_{k=1}^{n} \{ -\frac{c_{1}}{2} (\left | \gamma _{i,k} \right |-\kappa _{i,k} )^{2p}f_{i,k}- \frac{c_{2}}{2} (\left | \gamma _{i,k} \right |-\kappa _{i,k} )^{2q}f_{i,k}\}\nonumber\\+&\sum_{i=1}^{N}c_{3}\{-\left \| e_{i} \right \| ^2 -\left \| e_{i} \right \| ^{2p} -\left \| e_{i} \right \| ^{2q}  +\left \| e_{i} \right \| ^{2p}+c_{3}\left \| e_{i} \right \| ^{2q} \}\nonumber\\ +&\sum_{i=1}^{N}\sum_{k=1}^{n}c_{4}\{-(\frac{1}{2r_{i,k}}\tilde{\varphi}^2_{i,k})-(\frac{1}{2r_{i,k}}\tilde{\varphi}^2_{i,k})^p  -(\frac{1}{2r_{i,k}}\tilde{\varphi}^2_{i,k})^q \nonumber\\+&(\frac{1}{2r_{i,k}}\tilde{\varphi}^2_{i,k})^p  +(\frac{1}{2r_{i,k}}\tilde{\varphi}^2_{i,k})^q\}+\breve{\Pi}
\end{align}
where $ \breve{\Pi}=\sum_{i=1}^{N}(\sum_{k=1}^{n}\bar{m }\varsigma \varphi _{i,k} +0.2875\varepsilon _{i}+q_{0}),c_{1}=2a_{i,k},c_{2}=2b_{i,k},c_{3}=\varpi_{i},c_{4}=2\rho_{i,k}$. To satisfy the tracking control performance and make the systems convergence, the parameters' values should be selected carefully, and rules ought to be followed: $a_{i,k}>0,b_{i,k}>0,\varpi_{i}>0,r_{i,k}>0,\rho_{i,k}>0$.

Finally, we get
\begin{align}
\dot{V} \le -\beta _{1}V^p-\beta _{2}V^q+c\sum_{i=1}^{N}\left \| e_{i} \right \| ^{2p} +c(\sum_{i=1}^{N}\sum_{k=1}^{n} \frac{1}{2r_{i,k}}\tilde{\varphi} ^2 _{i,k})^p\nonumber+\Pi^*
\end{align}
where $c=\min\{2^pc_{1},2^qc_{2},c_{3},c_{4}\}$,$\beta_{1}=cN^{1-p},\beta_{2}=c,\Pi^*=\breve{\Pi}+2c(1-q)q^{\frac{q}{1-q}}$.
Then, $e_{i},\tilde{\varphi}_{i,k}$ with $ |e_{i}|\leq\theta_{1},|\tilde{\varphi}_{i,k}|\leq \theta_{2}$ are positive unknown constants, we have
\begin{align}
\dot{V} \le -&\beta _{1}V^p-\beta _{2}V^q+\bar{\Pi}
\end{align}
where $\bar{\Pi}=c\sum_{i=1}^{N}\left \| \theta_{1} \right \| ^{2p} +c(\sum_{i=1}^{N}\sum_{k=1}^{n} \frac{1}{2r_{i,k}}\theta_{2} ^2 )^p+\Pi^* .$

Based on Lemma 1 in \cite{ref39}, we get that every signal of MASs (1) can approach the compact set. Additionally, we have $\left \| y_{i}-y_{0} \right \|\le 2(\frac{\bar{\Pi} }{(1-\mho )\beta _{1}} ) $, which shows that the error between followers and the output signal are bounded.

Then, we construct a newly specified error variable in order to demonstrate that the system can guarantee that each tracking consensus error could converge to a predetermined interval asymptotically.
\begin{align}
\gamma_{i}=(|\gamma_{i,1}|-\kappa_{i,1})^pf_{i,1}.
\end{align}

Invoking Lemma 1 in \cite{ref39} and (52), we have $\dot{V}<0$. And the derivation of $\gamma_{i}$ is denoted by $\dot{\gamma}_{i}=p(|\gamma_{i,1}|-\kappa_{i,1})^{p-1}sg_{i,1}(\gamma_{i,1})f_{i,1}\dot{\gamma}_{i} $, which satisfies $\dot{\gamma}_{i}$ is bounded. We can get: $\int_{0}^{\infty } \gamma^2_{i} dt \le \frac{1}{a_{i,1}}V_{i,n} (0)$.

Owing to Barbalat's lemma as well as the convergence of $\dot{\gamma}_{i}$, we get $\lim_{t \to T_{0}}|\gamma_{i}|=0 $. So $\lim_{t \to T_{0}}|\gamma_{i,1}|\leq \kappa_{i,1} $ . From (44), we can obtain:$
\dot{w} _{i}(t)=\dot{\alpha}  _{i,n}-\frac{\xi _{i}{A_{i,n}}' }{cosh^2(\frac{A_{i,n}\xi _{i}}{\varepsilon _{i}}) } .$

According to \cite{ref40}, we get
\begin{align}
\frac{\mathrm{d} \breve{\phi} _{i}(t)}{\mathrm{d} t} =\frac{\mathrm{d} (\breve{\phi} _{i}(t)\times\breve{\phi} _{i}(t))^\frac{1}{2}  }{\mathrm{d} t}=sign(\breve{\phi } _{i}(t)\dot{\breve{\phi}}  _{i} (t))\le \left | \dot{w} _{i}(t) \right |   \nonumber.
\end{align}

{\bf{Remark 5:}} Based on the above analysis, there is a positive constant $\delta$ satisfying $\left | \dot{w} _{i}(t) \right |\leq \delta$. Invoking (44) and (45), we get $\lim_{t \to t_{\beta +1}}\phi _{i(t)}=\xi ^*_{i}$ and $\phi _{i(t)}=0$. Additionally, $t^* \ge (\frac{\xi ^*_{i}}{\delta})$ must be satisfied as the lowest limit of the inter-execution intervals for $ \forall t \in [t_{\beta},t_{\beta+1})$, so the Zeno behavior could be removed successfully.

Combining (51) and Lemma 2, the MASs are fixed-time stable, and the setting time  is denoted in the following:
\begin{align}
T_{0}=\frac{1}{\eta\beta_{1}(1-p)}+ \frac{1}{\eta\beta_{2}(q-1)}
\end{align}
where $\bar{\Pi}<\min{(1-\eta)\beta_{1},(1-\eta)\beta_{2}},(\eta\in(0,1))$ and it can be decided in advance.

\section{SIMULATION RESULTS}
{\bf{Example:}} Through the simulation, we could show how effective our anticipated method is; We study the MASs with followers 1-4 and the leader 0. The communication digraph for MASs $G^*$ is denoted:
\begin{align}
A=\begin{bmatrix} 0 &0  &0  &0 \\  1& 0 &  0& 1\\  1& 0 &0  &0 \\  0& 1 & 0 &0\end{bmatrix},L=\begin{bmatrix} 0 &0  &0  &0 \\  -1& 1 &  0& -1\\  -1& 0 &1  &0 \\  0& -1 & 0 &0\end{bmatrix} \nonumber
\end{align}

Additionally, $D=diag\{1,0,0,1\}$.

Each follower's states information is described
\begin{align}
\dot{x} _{i,1}=&x_{i,2}+\frac{x_{i,1}}{1+x^2_{i,2}}\nonumber\\
\dot{x} _{i,2} =&\bar{u}_{i}+\breve{o}_{i}sin(x_{i,1}-x_{i,2})e^{-(x^2_{i,1}+x^4_{i,2})}, i=&1,2,3,4\nonumber
\end{align}
with $\breve{o}_{1}=\breve{o}_{2}=\breve{o}_{3}=\breve{o}_{4}=0.15$. The leader's signal is $y_{0}=3\sin(2t)$. By the event-triggered controller and update rules we designed, we can ensure that each agent follows the leader in a fixed time. NNs have 16 nodes with centers in the range of [-0.5,0.5] in simulation example. The remaining parameters are defined in Table \ref{table:1}.

\begin{table}[h]
\caption{PARAMETERS OF CONTROL SCHEME.\label{table:1}}
\newcolumntype{C}{>{\centering\arraybackslash}X}
\scalebox{1.1}{ 
\begin{tabular}{|c c c c|}
 \hline
  $x_{0}(0)=0$ & $x_{1,1}(0)=0.2$&$x_{1,2}(0)=0.3$ &$x_{2,1}(0)=0.2$ \\
   $x_{2,2}(0)=0.3$ & $x_{3,1}(0)=0.2$&$x_{3,2}(0)=0.5$&$x_{4,1}(0)=0.3$ \\
 $x_{4,2}(0)=0.2$&$\hat{x}_{1,1}(0)=0.2$&$\hat{x}_{1,2}(0)=0.3$&$\hat{x}_{2,1}(0)=0.2$\\
 $\hat{x}_{2,2}(0)=0.3$&$\hat{x}_{3,1}(0)=0.2$&$\hat{x}_{3,2}(0)=0.5$&$\hat{x}_{4,1}(0)=0.3$\\
 $\hat{x}_{4,2}(0)=0.2$&$\hat{\varphi}_{1}(0)=0.1$&$\hat{\varphi}_{2}(0)=0.1$&$\hat{\varphi}_{3}(0)=0.1$\\
 $\hat{\varphi}_{4}(0)=0.1$&$a_{1,1}=15$&$a_{2,1}=15$&$a_{3,1}=15$\\
$a_{4,1}=15$&$a_{1,2}=2$&$a_{2,2}=2$&$a_{3,2}=2$ \\
$a_{4,2}=2$&$b_{1,1}=35$&$b_{2,1}=35$&$b_{3,1}=35$\\
$b_{4,1}=35$&$b_{1,2}=5$&$b_{2,2}=5$&$b_{3,2}=5$\\
$b_{4,2}=5$&$s_{i}=10$&$r_{i,1}=1$&$r_{i,2}=1$\\
$\kappa_{i,1}=0.6$&$\kappa_{i,2}=0.8$&$p=2$&$q=0.5$\\
$\xi_{1}=5.5$&$\xi_{2}=4.5$&$\xi_{3}=4.5$&$\xi_{4}=5$\\
$\xi^*_{1}=5$&$\xi^*_{2}=4$&$\xi^*_{3}=4$&$\xi^*_{4}=4.5$\\
$\varepsilon_{1}=25$&$\varepsilon_{2}=25$&$\varepsilon_{3}=25$&$\varepsilon_{4}=25$\\
$\mu_{i,1}=-15$&$\mu_{i,2}=-80$&$\rho_{i,1}=1$&$\rho_{i,2}=1$\\[1ex]
 \hline
\end{tabular}
}
\end{table}

\begin{figure}[h]
\centering
\includegraphics[height=6.5cm,width=9cm]{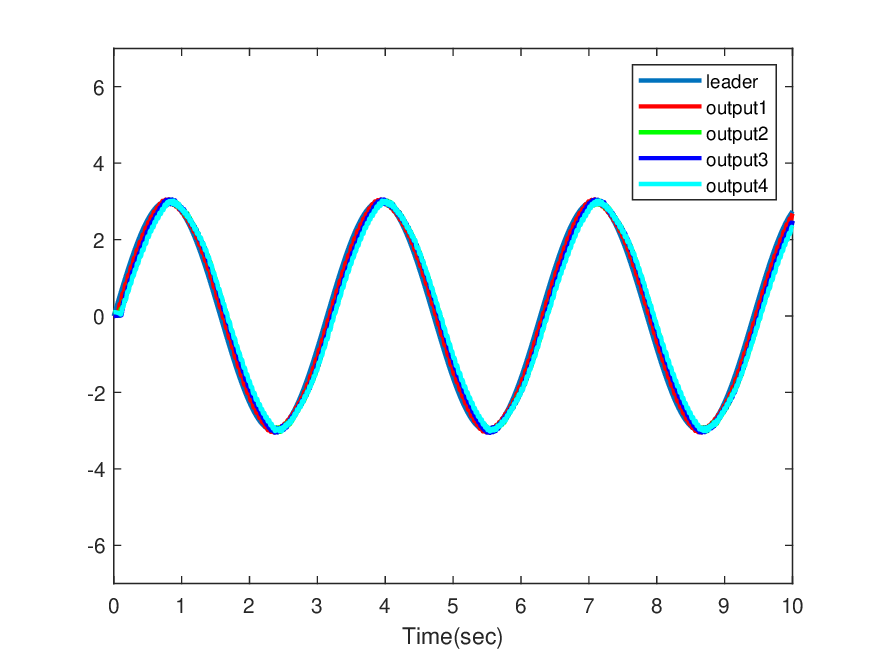}
\captionsetup{singlelinecheck=off}
\caption{Tracking Curves.}
\label{figure 2}
\end{figure}

\begin{figure}[h]
\centering
\includegraphics[height=6.5cm,width=9cm]{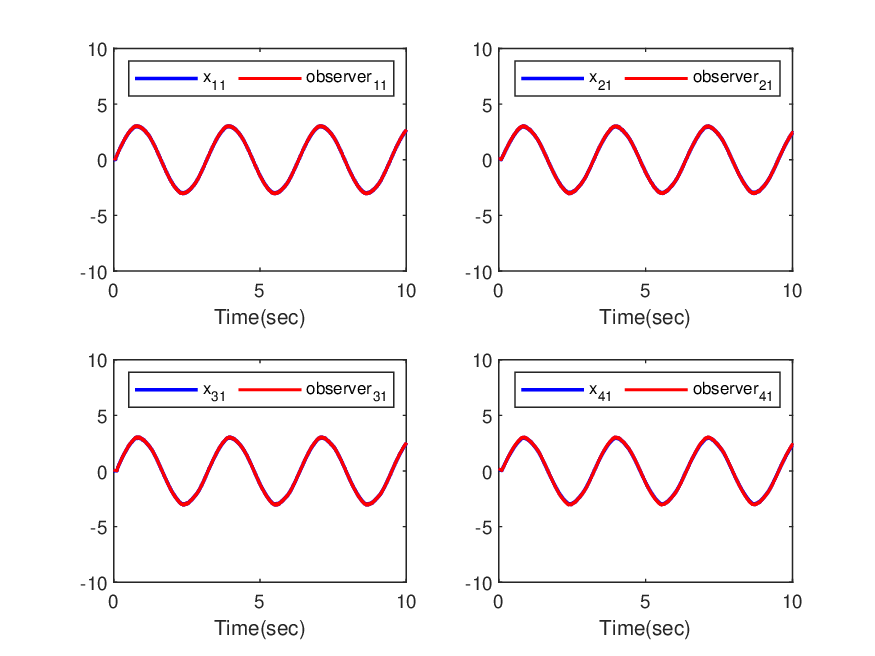}
\captionsetup{singlelinecheck=off}
\caption{States $x_{i,1}$ and estimated values $\hat{x}_{i,1}, i=1,2,3,4$.}
\label{figure 3}
\end{figure}

\begin{figure}[h]
\centering
\includegraphics[height=6.5cm,width=9cm]{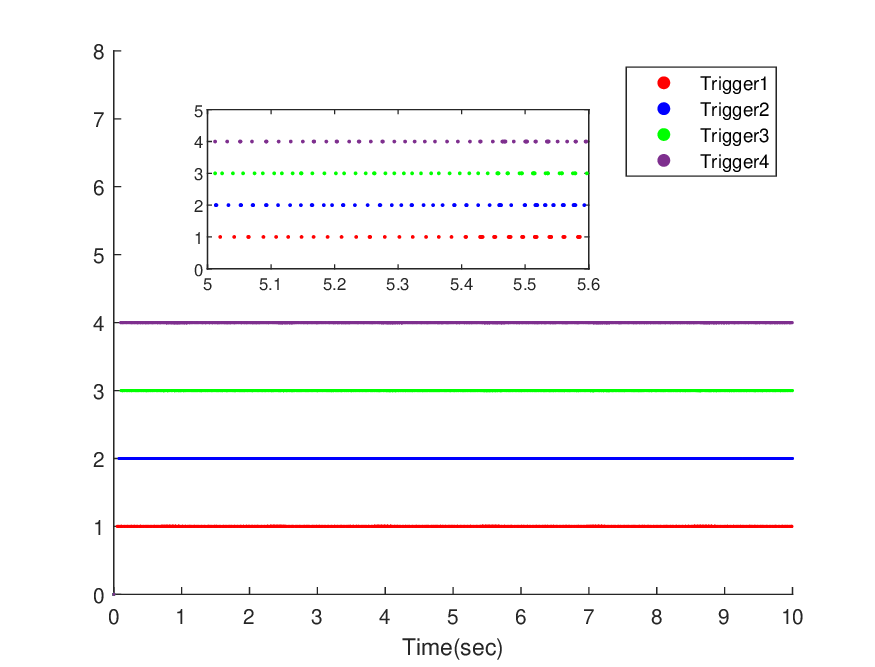}
\captionsetup{singlelinecheck=off}
\caption{Triggering event.}
\label{figure 6}
\end{figure}

Figure \ref{figure 2}, \ref{figure 3}, and \ref{figure 6} state the simulation results. The output curves are shown in figure\ref{figure 2}. Figure \ref{figure 3} denotes the 4 follower states and the observer state respectively. It apparently bounded. Figure \ref{figure 6} Shows the triggering time. Apparently, the controller is not triggered all the time to save the communications resources. Each follower's triggering numbers are 738, 943, 750, and 872, respectively. Figure \ref{figure 2}-\ref{figure 6} show all the signals are bounded.

\section{CONCLUSION}
The event-triggered observer-based fixed-time consensus problem for uncertain nonlinear MASs is examined in our work. We use fixed-time control laws to ensure efficient convergence and independence from initial values, and a state observer to approximate unknown states. An adaptive event-triggered consensus control strategy is designed to save resources and prevent Zeno behavior. RBF NNs are used to handle complex functions, ensuring stability in fixed time. Simulation results demonstrate the effectiveness of our method. In addition, our future research will focus on the consideration of physically more complex models, applied for real complex systems, in order to improve the practicality of control methods.

\end{document}